\documentstyle[epsfig,a4,12pt]{article}
\def\tou#1{{\lower1.2ex\hbox{$\longrightarrow$}\atop
        {\lower-.7ex\hbox{$\scriptscriptstyle #1 $}}}}
\def\lsim{{\lower1.2ex\hbox{$<$}\atop
        {\lower-.7ex\hbox{$\sim$}}}}
\def\gsim{{\lower1.2ex\hbox{$>$}\atop
        {\lower-.7ex\hbox{$\sim$}}}}

\def\be{\begin{equation}}
\def\ee{\end{equation}}

\begin{document}

\begin{titlepage}

\rightline {Si-97-03 \  \  \  \   }
\vspace*{2.truecm} 

\centerline{\Large \bf Monte Carlo Measurement of }
\vskip 0.6truecm
\centerline{\Large \bf the Global Persistence Exponent\footnote
{Work supported in part by the Deutsche Forschungsgemeinschaft;
 DFG~Schu 95/9-1}}
\vskip 0.6truecm

\vskip 2.0truecm
\centerline{\bf L. Sch\"ulke and B. Zheng}
\vskip 0.2truecm
\centerline{Universit\"at -- GH Siegen, D -- 57068 Siegen, 
Germany}

\vskip 2.5truecm

\abstract{
The scaling behaviour of the persistence
probability in the critical 
dynamics is investigated with both the heat-bath and
the Metropolis algorithm for the
two--dimensional Ising model and Potts model.
 Special attention is drawn to
the dependence on the initial magnetization.
The global persistence exponent is measured. 
Universality is confirmed.
}

\vspace{0.5cm}


\vspace{0.3cm}


\end{titlepage}

For long a set of two static exponents $\beta$, $\nu$
and one dynamic exponent $z$ has been used to characterize the
scaling behaviour of critical dynamics.
This is sufficient when the dynamic systems
almost reach the equilibrium. When the dynamic systems
are far from the equilibrium, it was believed that
the behaviour of the dynamic systems essentially
depends on the microscopic details and there exists no 
universal scaling behaviour. 
However, it has recently been discovered that 
for a dynamic process, in which a magnetic system initially at 
very
high temperature is suddenly quenched to the critical
temperature and then released to the dynamic evolution of model 
A,
already in the 
{\it macroscopic short-time regime} emerges {\it universal} 
scaling
behaviour.
A new independent exponent should be introduced
to describe the dependence of the scaling behaviour
on the initial magnetization
\cite {jan89,hus89,hum91}. 
Furthermore, if a non-zero but small magnetization 
is given to the initial state,
at the macroscopic early time
the magnetization
surprisingly undergoes
{\it a critical initial increase} 
\cite {jan89,li94,gra95,sch95,oka97a,oka97}
\begin{equation}
M(t) \sim m_0 \, t^ \theta,
\label{e10}
\end{equation}
where $\theta$ is the new critical exponent which is
 related to the dimension $x_0$ of
the initial magnetization  $m_0$ by
$\theta=(x_0-\beta/\nu)/z$.

Are there any more critical exponents in the dynamic system
described above?
Very recently, it was argued that the global persistence 
exponent
of the magnetization is another new independent critical
exponent \cite {maj96a}. 
This relies on the observation that the dynamic evolution
of the magnetization is {\it not a Markovian process}.
If a non-zero magnetization
is given to the initial state at very high temperature,
the time-dependent magnetization keeps its
sign at the time $t$ with a probability $p(t)$.
In the limit of  {\it zero} initial magnetization, the 
probability
$p(t)$ decays by a power law with respect to the time $t$
\begin{equation}
p(t) \sim t^ {-\theta_1}.
\label{e20}
\end{equation}
Here $\theta_1$ is called the global persistence exponent.
The exponent $\theta_1$ has been calculated 
perturbatively for the $O(N)$
vector model and numerically determined
for the two--dimensional Ising model 
with the heat-bath algorithm \cite {maj96a}.
In reference \cite {maj96a}, the power law behaviour 
in Eq.~(\ref{e20}) was not clearly observed since 
the lattice size was not big enough.
Only a combination of the exponents $\theta_1 z$
has been obtained from finite size scaling.
The result carried a relatively big error.
Stauffer has performed simulations with the heat-bath algorithm
for bigger lattice
sizes and estimated the exponent $\theta_1$ from
the power law behaviour 
 for the two-- and three--dimensional
Ising model \cite {sta96}. For the two--dimensional Ising model
with lattice size $L=300$, the result is $\theta_1=0.225$.
The error is probably somewhat below
$0.01$ \cite {sta96}.
If the time evolution of the magnetization is
a Markovian process, a scaling law which relates
both the exponent $\theta$ and $\theta_1$
 \begin{equation}
\theta_1 = \alpha_1 \equiv - \theta + (1-\beta/\nu)/z
\label{e30}
\end{equation}
can be deduced. 
It has been argued at the two-loop level of an $\epsilon$--expansion
that this scaling law is violated
 \cite {maj96a}.
It is interesting to investigate numerically 
how big the violation of the scaling law is.
Due to the relative big errors for the exponent $\theta_1$
obtained in references \cite {sta96,maj96a}, the results
are not completely 
satisfactory\footnote {When the revised version of this paper 
was completed, 
the two loop contribution has been explicitly
calculated \cite {oer97}.}.
 
In this letter, we present systematic Monte Carlo simulations
for the scaling behaviour of the persistence probability
of the magnetization. Special attention is drawn
to the dependence of the scaling behaviour on the initial
magnetization. Calculations are extended
to the two--dimensional Potts model. Further, to confirm universality,
all the simulations
are carried out with both the heat-bath and Metropolis algorithm.

At first, let us concentrate on the two--dimensional Ising model.
In reference \cite {maj96a}, the initial state is simply generated
{\it randomly} --- with equal probability of spins up and down. 
Since the lattice size in practical simulations
is finite, it is almost always the case that a finite non-zero 
initial
magnetization will be generated
in each initial configuration. 
Then the system is released to evolve at the critical 
temperature
with the heat-bath algorithm.  
We update the system until the magnetization change its sign.
Repeating this procedure many times with different initial 
configurations
and random numbers, we can measure the global
persistence probability $p(t)$.
If the lattice is sufficient large, $p(t)$ should decay
by the power law given in Eq.(\ref{e20}).
As was observed in reference \cite {maj96a}, however,
the lattice size $L=128$ is not sufficiently big
and the power law is not clearly observed due to the
finite size effect.

Here we should point out that in this dynamic system
there are two kinds of finite size effects.
One is the normal finite size effect which 
takes place in a time scale $ t_L \sim L ^ z$.
Whenever the system evolves into this time regime,
$p(t)$ will decay by an exponential law
$\exp(-t/t_L)$ rather than a power law.
For a lattice size as $L=128$ or bigger,
the time scale $ t_L $ is quite big. Later we will see that
in the short-time regime this effect is not prominent.
Another kind of finite size effect is an extra  
 finite size effect from
 the initial configurations.
When the lattice size is finite, each initial configuration
generated {\it randomly} as described above has a finite 
non-zero
initial magnetization. This non-zero initial magnetization
tends to zero only in the limit of infinite lattice size.
Such a finite non-zero initial magnetization
 would modify the power law behaviour
 in Eq.(\ref{e20}) even in the short-time regime of
 the dynamic evolution.   

 In order to handle the extra finite size effect 
 efficiently and to discuss the dependence of the scaling behaviour 
on the initial magnetization,
 we introduce {\it a sharp preparation technique} in 
 preparing the initial configurations: first the initial configuration
 is generated {\it randomly}; then we select 
 a site randomly on the lattice and flip the spin if the updated
 magnetization comes nearer to a pre-fixed small 
 value $m_0$; we continue this procedure until
 the initial magnetization reaches the value $m_0$.
 By this technique, we can obtain any fixed value
for the initial magnetization.

\begin{figure}[t]\centering
\epsfysize=12cm
\epsfclipoff
\fboxsep=0pt
\setlength{\unitlength}{1cm}
\begin{picture}(13.6,12)(0,0)
\put(1.9,11.0){\makebox(0,0){1}}
\put(1.9,4.0){\makebox(0,0){0.1}}
\put(1.2,8.0){\makebox(0,0){$p(t)$}}
\put(11.8,1.2){\makebox(0,0){$t$}}
\put(9.7,9.0){\makebox(0,0){$L=128,  m_0=0.01$}}
\put(6.0,5.5){\makebox(0,0){$m_0=0.001, 0.0005$}}
\put(0,0){{\epsffile{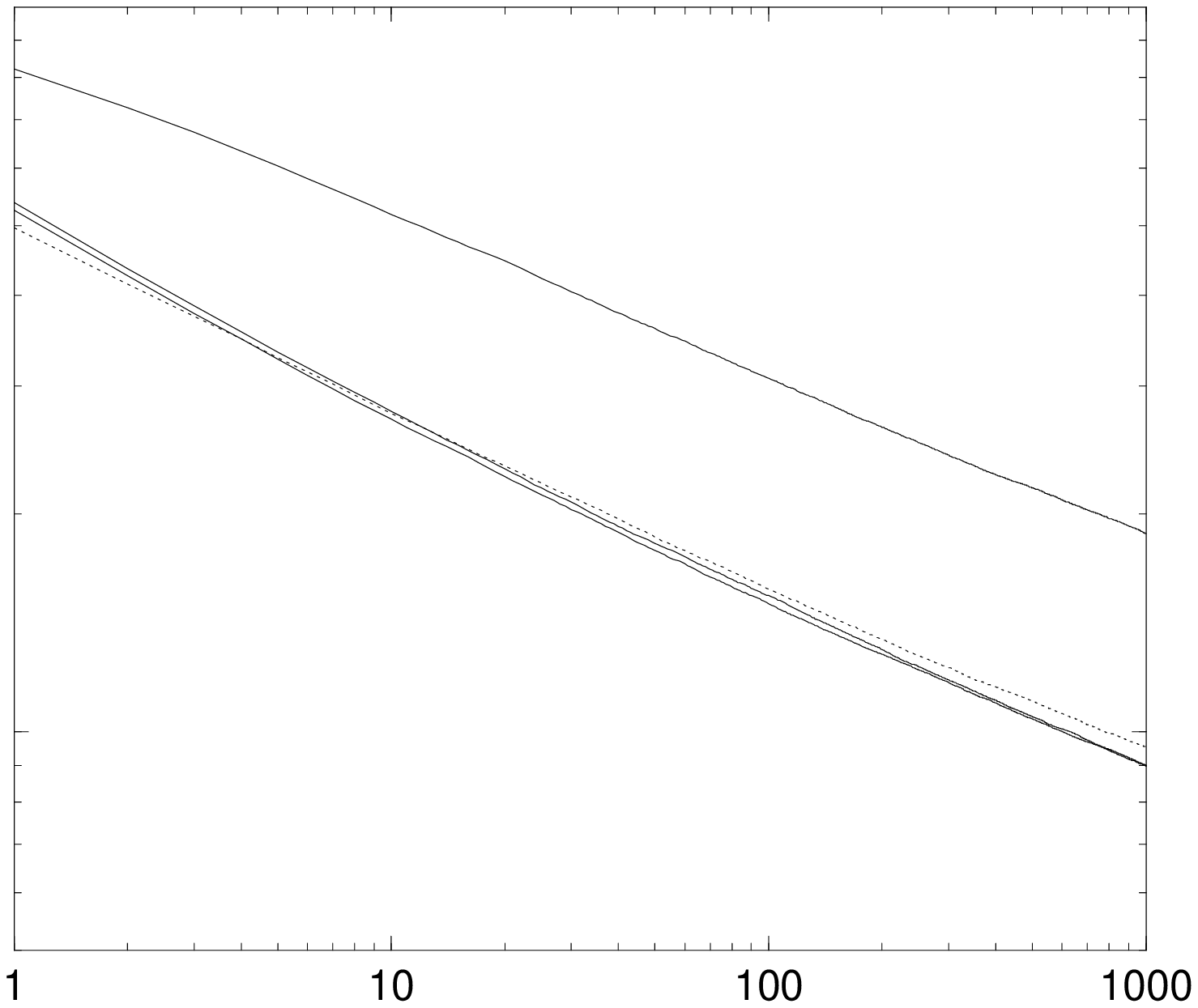}}}
\end{picture}
\caption{ The persistence probability $p(t)$ for the Ising model
with the heat-bath algorithm. The lattice size is $L=128$.
Solid lines are the results with the sharp preparation of the 
initial
magnetization. The initial magnetization
$m_0$ is $0.01$, $0.001$ and $0.0005$ (from above).
The dotted line is that without the sharp preparation technique.
}
\label{f1}
\end{figure}

 In Fig.~\ref{f1}, the persistence probability $p(t)$
 is displayed with solid lines
 in double-log scale for the lattice size
 $L=128$ and different
 initial magnetization $m_0$. For comparison, 
 the result without the sharp preparation of the initial
 magnetization is also plotted  with a dotted line.
Samples of independent initial configurations are
$20\ 000$, $40\ 000$ and $60\ 000$ for $m_0=0.01$, $0.001$ and
$0.0005$ respectively and that without the sharp preparation
of the initial magnetization is $40\ 000$.
 From the figure, we see clearly that the behaviour
 of $p(t)$ apparently depends on the initial magnetization.
 When the initial magnetization tends to zero,
 however, it convergences. The difference between
 the curves with $m_0=0.001$ and $m_0=0.0005$ is already 
 very small. The curve without the sharp preparation
 technique is a certain combination of those with
 different $m_0$. For the lattice size
 $L=128$, the curves with small $m_0$ do also not yet
 present perfect power law behaviour even though
 it
 is improved compared with that without the sharp preparation
technique.

\begin{figure}[t]\centering
\epsfysize=12cm
\epsfclipoff
\fboxsep=0pt
\setlength{\unitlength}{1cm}
\begin{picture}(13.6,12)(0,0)
\put(1.9,11.0){\makebox(0,0){1}}
\put(1.9,4.0){\makebox(0,0){0.1}}
\put(1.2,8.0){\makebox(0,0){$p(t)$}}
\put(11.8,1.2){\makebox(0,0){$t$}}
\put(9.7,9.0){\makebox(0,0){$m_0=0.0005$}}
\put(0,0){{\epsffile{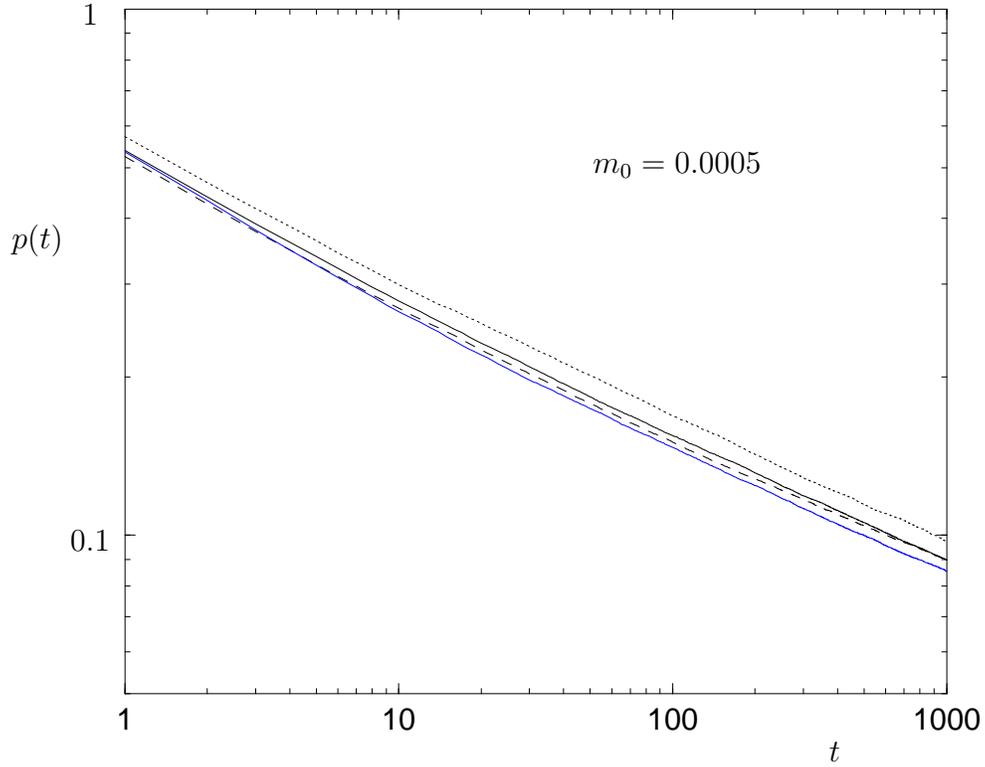}}}
\end{picture}
\caption{ The persistence probability $p(t)$ for the Ising model
with the heat-bath algorithm and the Metropolis algorithm.
The sharp preparation technique is adopted with $m_0=0.0005$.
The dashed line, upper solid line and the dotted line
are the results of the heat-bath algorithm for
the lattice sizes $L=128$, $256$ and $512$ respectively.
The lower solid line is that of the Metropolis algorithm
for the lattice size $L=256$. 
}
\label{f2}
\end{figure}

 In Fig.~\ref{f2}, the persistence probability $p(t)$
 is displayed in double-log scale for different lattice
 sizes. The sharp preparation technique is adopted.
 $m_0$ is taken to be $0.0005$. The dashed line, upper
 solid line and the dotted line correspond to the
 lattice size $L=128$, $256$ and $512$ respectively.
 Samples of the independent initial configurations are
$60\ 000$ and $28\ 000$ for the lattice sizes 
$L=256$ and $512$. 
In the figure, it can be seen that 
 after a microscopic time scale $t_{mic} \sim 50$,
 the power law decay is observed for bigger lattices.
 If we measure the slopes of these three curves from
 a time interval $[50,1000]$, the resulting
 exponent $\theta_1$ is $0.224(2)$, $0.238(3)$ 
 and $0.238(6)$ respectively. 
The errors are estimated by dividing the data
into three or four groups.
 As pointed out above,
 there is still some extra finite size effect
 for the lattice size $L=128$, however, the lattice size
 $L=256$ and $L=512$ give the same results within
 the statistical errors. The extra finite size effect
 is already negligible small.  
 
 In Fig.~\ref{f2},
 we can also see that the normal finite size effect has not yet
 appeared up to the evolution time $t=1000$. 
 The normal finite size effect is characterized
 by an exponential decay of $p(t)$, which is faster 
 than the power law decay. However,  for the smaller 
 lattice size,  e.g. 
 $L=128$, the practically measured exponent is even smaller.
 Such a tendency can also be seen clearly in the figure.

Are the power law behaviour of $p(t)$  and the
critical exponent $\theta_1$ universal?
In order to check this point, we have also performed 
simulations with the Metropolis algorithm.
For smaller lattice sizes, the behaviour of $p(t)$
with the heat-bath and the Metropolis algorithm is
somehow different. When the lattice is sufficient
large, however, both algorithms present almost the same
power law behaviour. The result of $p(t)$
for the Metropolis
algorithm with lattice size $L=256$ and $m_0=0.0005$
is displayed with the lower solid line in Fig.~\ref{f2}. 
The corresponding
exponent is $\theta_1 = 0.236(3)$.
Within the statistical errors, the result is consistent
with that of the heat-bath algorithm. Universality is confirmed.

\begin{figure}[t]\centering
\epsfysize=12cm
\epsfclipoff
\fboxsep=0pt
\setlength{\unitlength}{1cm}
\begin{picture}(13.6,12)(0,0)
\put(1.9,11.0){\makebox(0,0){1}}
\put(1.9,4.0){\makebox(0,0){0.1}}
\put(1.2,8.0){\makebox(0,0){$p(t)$}}
\put(11.8,1.2){\makebox(0,0){$t$}}
\put(9.7,7.0){\makebox(0,0){$L=256,  m_0=0.001$}}
\put(6.5,5.5){\makebox(0,0){$m_0=0.0005$}}
\put(0,0){{\epsffile{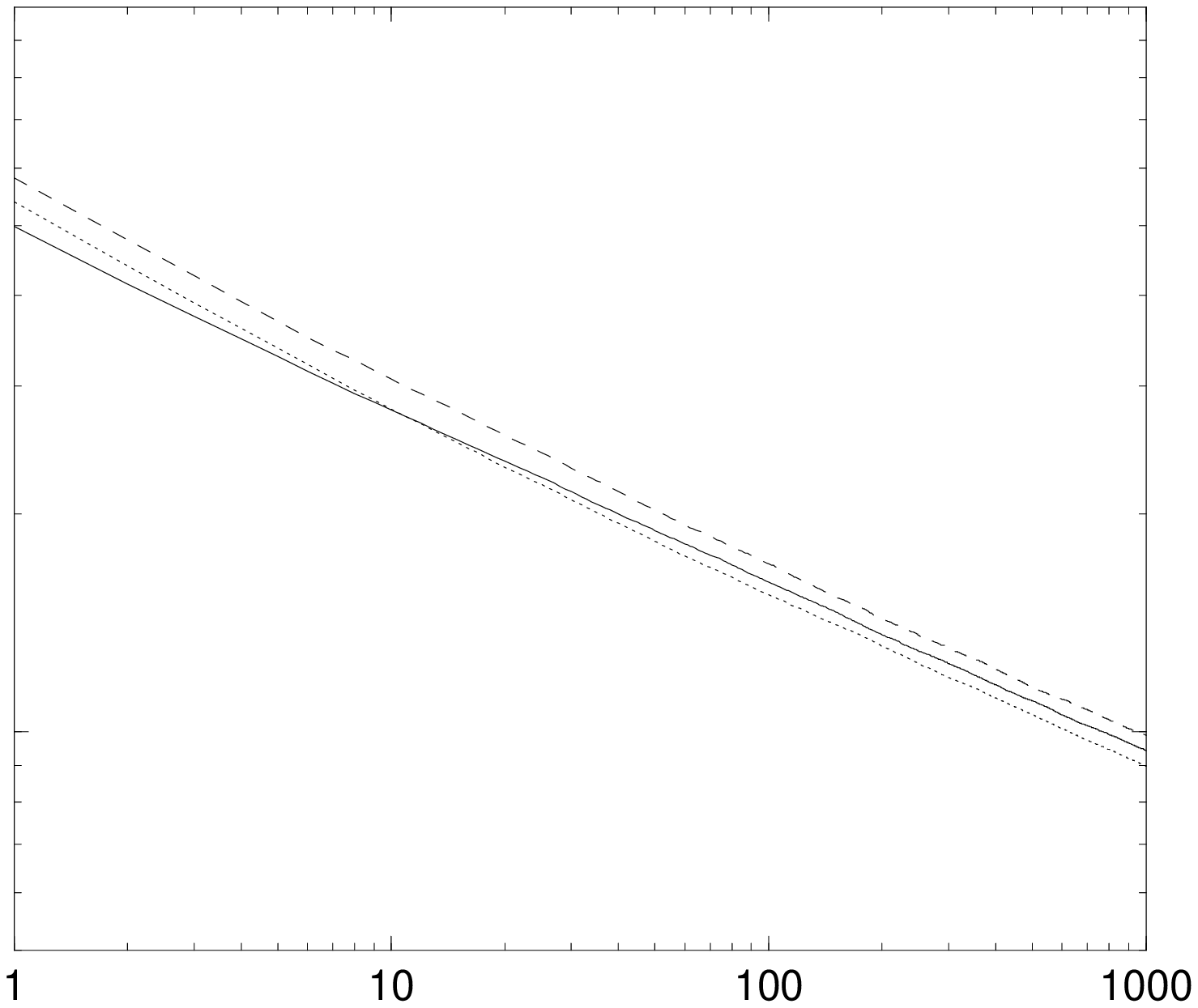}}}
\end{picture}
\caption{The persistence probability $p(t)$ for the Ising model
with the heat-bath algorithm.   The lattice size is $L=256$.
The dotted line and the dashed line are the results for
$m_0=0.0005$ and $0.001$ respectively, while the solid line 
shows the result for the initial magnetization not sharply 
prepared.
}
\label{f3}
\end{figure}

To have more confidence on our measurements, we have also
performed simulations with the heat-bath algorithm
 for the lattice size $L=256$
with $m_0=0.001$ and 
with initial magnetization not sharply prepared. 
The results are displayed 
with a dashed line and a solid line in Fig.~\ref{f3}.
The measured exponents are $0.237(5)$ and $0.233(5)$
respectively. The result of $m_0=0.001$ is 
the same as that of $m_0=0.0005$ within the errors. The result
without the sharp preparation technique
is slightly smaller, but the extra finite size effect
is for L=256 already not so big.
Our result $\theta_1=0.233(5)$ without the sharply prepared
initial magnetization, i.e. with a {\it random} initial
state, is slightly different from what Stauffer obtained,
$\theta_1=0.225(10)$. But the difference is
 still within the stattistical errors.
 For comparison, the result 
for $L=256$ and $m_0=0.0005$ is also plotted with
the dotted line in Fig.~\ref{f3}.

All the results for the Ising model have been summarized
in Table~\ref{t1} and Table~\ref{t2}. 
 Samples of the independent initial configurations are
from $28\ 000$ to $60\ 000$, depending on the lattice size $L$
and the initial magnetization $m_0$. Errors are estimated by 
dividing 
the samples into three or four groups.
Compared with the result obtained
from the finite size scaling in reference \cite {maj96a},
our results are more accurate. 

\begin{figure}[t]\centering
\epsfysize=12cm
\epsfclipoff
\fboxsep=0pt
\setlength{\unitlength}{1cm}
\begin{picture}(13.6,12)(0,0)
\put(1.9,11.0){\makebox(0,0){1}}
\put(1.9,5.1){\makebox(0,0){0.1}}
\put(1.2,8.0){\makebox(0,0){$p(t)$}}
\put(11.8,1.2){\makebox(0,0){$t$}}
\put(9.7,7.0){\makebox(0,0){$heat-bath$}}
\put(6.5,5.5){\makebox(0,0){$Metropolis$}}
\put(0,0){{\epsffile{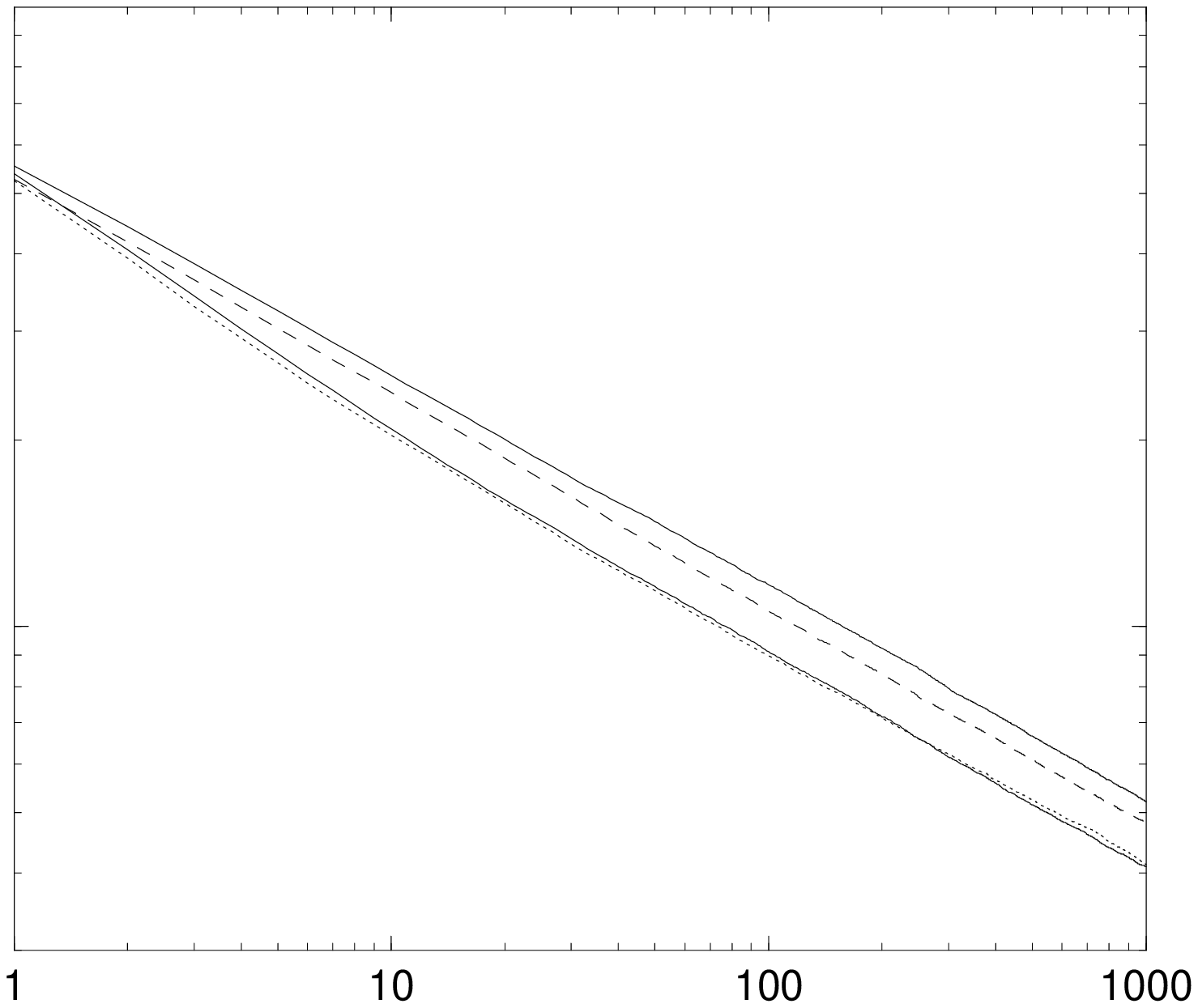}}}
\end{picture}
\caption{ The persistence probability $p(t)$ for the Potts model
with the heat-bath algorithm and the Metropolis algorithm.
The sharp preparation technique is adopted and $m_0=0.0005$.
The dashed line and the upper solid line are the results of the
heat-bath algorithm for the lattice sizes $L=144$ and $L=288$
respectively. The dotted line and the lower solid line are the 
results of the
Metropolis algorithm for the lattice sizes $L=144$ and $L=288$.
}
\label{f4}
\end{figure}

Encouraged by the success for the Ising model,
we have carried out similar simulations 
for the two--dimensional Potts model.
In Fig.~\ref{f4}, $p(t)$ has been plotted in double-log
scale with $m_0=0.0005$.
The dashed line and the upper solid line are the result
with the heat-bath algorithm for lattices $L=144$ and $288$, 
while
the dotted line and the lower solid line
are those with the Metropolis algorithm.
Slightly different from the case of the Ising model,
for the lattice size $L=144$ the extra finite size effect
is already not so big. For example, with the heat-bath algorithm
the measured exponent $\theta_1$ is $0.344(6)$ and $0.350(1)$
for the lattice sizes $L=144$ and $L=288$ respectively.
Within the statistical errors, they cover each other.
In Table~\ref{t2}, the exponent $\theta_1$ measured for the 
Potts model
for the lattice size $L=288$ with both the heat-bath 
and the Metropolis algorithms is also given.
The statistics is $60\ 000$.

How big is the violation of the scaling law in Eq.(\ref{e30})
in the numerical sense?
The static critical exponent $\beta$ and $\nu$ 
for the two--dimensional Ising and Potts model
are exactly known, while
the dynamic exponent $\theta$ and $z$ have numerically
 been measured from the short-time dynamics \cite {oka97a,gra95}.
The value of $\alpha_1 =  - \theta + (1-\beta/\nu)/z$
 for the Ising model and the Potts model
with both the heat-bath and the Metropolis algorithm
are listed in Table~\ref{t2}. These results
are taken from reference \cite {oka97a}.
Apparently the exponent $\theta_1$ is not equal to
the exponent $\alpha_1$.
The difference is about $10$ percent.
For example, for the Ising model with the heat-bath algorithm
$\theta_1=0.238(3)$, while $\alpha_1=0.215(1)$.

Here we should mention that in the estimate of
the exponent $\alpha_1$ both exponents $\theta$ and $z$
are involved. The numerical measurements of $\theta$
are somehow satisfactory \cite {oka97a,gra95}.
The value of $\theta$ for the Potts model is also
relatively small and will not induce a big error
for $\alpha_1$. However, the exponent $z$ is 
problematic, at least for the Ising model.
Values ranging from $z=2.13$ to $z=2.17$ can be found in the
literature \cite {tan87,lan88,ito93,li95,li96,gra95,nig96}.
Fortunately, the value of $\alpha_1$ is not so sensitive
to the exponent $z$. For example, in reference \cite {gra95}
we find $\theta=0.191(3)$ and $z=2.172(6)$
for the Ising model with the heat-bath algorithm.
 Then one obtains $\alpha_1=0.212(3)$.
It is slightly smaller than that given in Table~\ref{t2},
but still consistent within the statistical errors.

In reference \cite {maj96a}, a combination
of the exponents $\theta_1 z = 0.505(20)$
 was obtained from finite size scaling.
This leads to the values 
$\theta_1=0.233(9)$ and $\theta_1=0.234(9)$
for $z=2.172(6)$ and $z=2.155(3)$ respectively.
These values are smaller compared with those
in Table~\ref{t2}, and 
taking into account the bigger error, the difference
between the exponent $\theta_1$ obtained here
 and $\alpha_1$ given in Table~\ref{t2}
is less prominent. --
In reference~\cite {sta96} the result $\theta_1=0.225$
with an error around $0.01$ is given. Here 
the values for $\theta_1$ and $\alpha_1$
nearly cover each other.

Finally, we investigate the general
scaling behaviour for the persistence probability
$p(t)$ in case the effect of the initial magnetization 
can not be neglected. Let us assume a finite
size scaling form 
\begin{equation}
p(t,L,m_{0})=t^{-\theta_1}
f(t^{1/z} L^{-1}, t^{x_0/z} m_0).
\label{e40}
\end{equation}
with $x_0$ being the scaling dimension of the initial
magnetization $m_0$. 
From the time evolution of the magnetization or 
the auto-correlation,
the value of $x_0$ for the Ising and Potts 
model has previously been determined 
directly \cite {li94} or indirectly through its relation
with the exponent $\theta$, i.e. $x_0=\theta z +\beta/\nu$
\cite {sch95,oka97a}.

\begin{figure}[t]\centering
\epsfysize=12cm
\epsfclipoff
\fboxsep=0pt
\setlength{\unitlength}{1cm}
\begin{picture}(13.6,12)(0,0)
\put(1.2,8.5){\makebox(0,0){$p(t)$}}
\put(10.8,1.2){\makebox(0,0){$t$}}
\put(0,0){{\epsffile{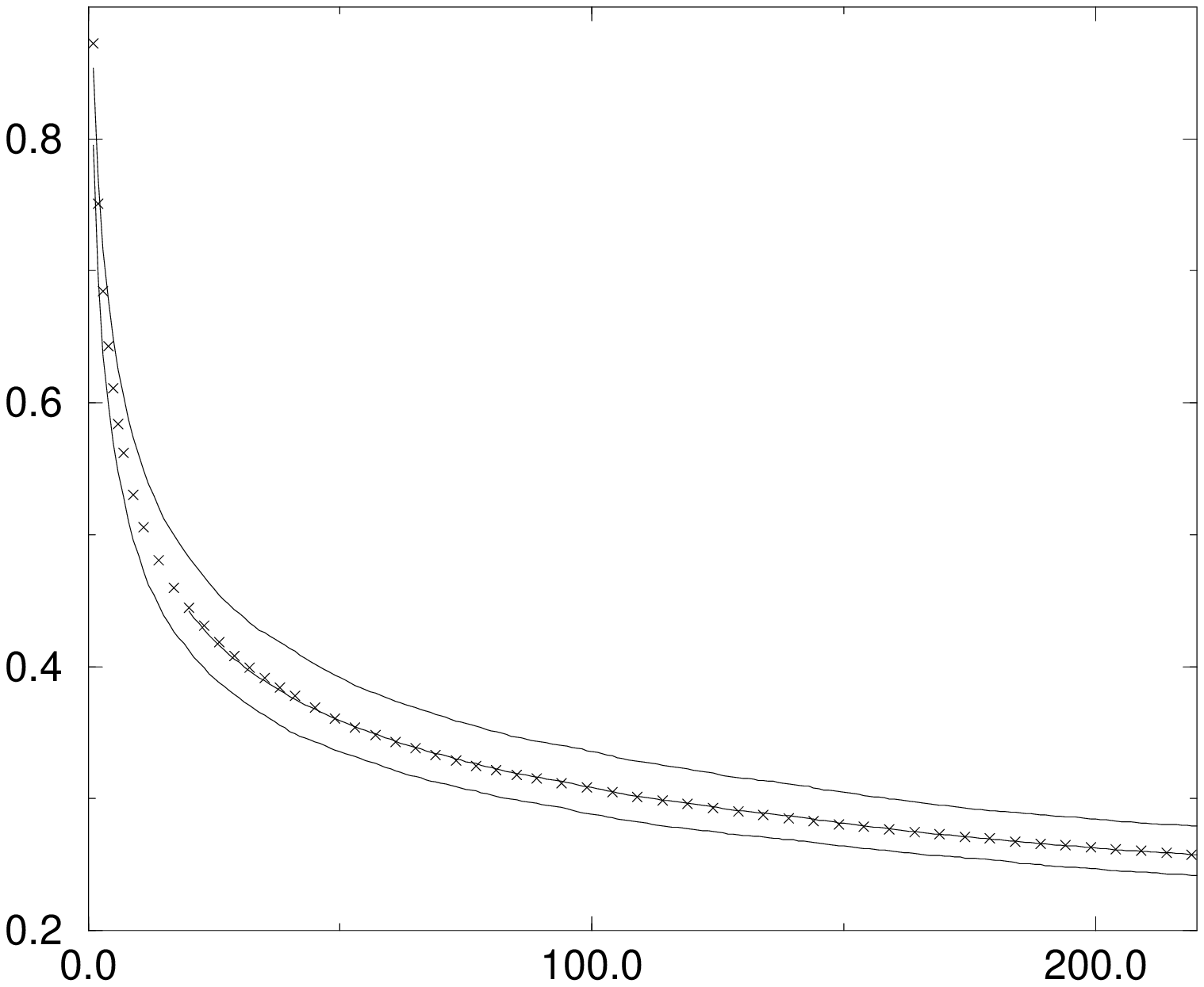}}}
\end{picture}
\caption{The scaling plot of the persistence probability
$p(t)$ for the two--dimensional Ising model with 
the heat-bath algorithm. The lower and upper solid lines are
the data for $L_2=128$ with $m_0=0.0090$ and 
$m_0=0.0115$ respectively.
The crosses are those
for $L_1=256$ with
$m_{01}=0.0050$.
The solid line fitted to the crosses corresponds
to $L_2=128$ and $m_{02}=0.0101(1)$.
}
\label{f5}
\end{figure}

Now, we intend to determine the scaling dimension $x_0$
from the scaling form in Eq. (\ref {e40}).
In order to do this, we take the two--dimensional
Ising model as an example and perform a simulation
for the lattice size $L_1=256$ with the initial magnetization
$m_{01}=0.0050$. If the scaling form in Eq. (\ref {e40}) holds,
one should be able to find an initial
magnetization $m_{02}$ for the lattice size $L_2=128$
such that the curves of $p(t)$ for both lattice sizes
collaps. From the ratio of $m_{01}$ and $m_{02}$,
one may estimate $x_0$. Practically we have performed
  simulations for $L_2=128$
with two values of the initial magnetization,
 $m_0=0.0090$ and $m_0=0.0115$.
By linear extrapolation, we can obtain data
with the initial magnetization
between these two values of $m_0$.
We search a curve which has a best fit to the curve for $L_1=256$
and then determine $m_{02}$. In Fig.~\ref {f5},
such a scaling plot is displayed.
The lower and upper
solid lines are the persistence probability 
for $L_2=128$ with $m_{0}=0.0090$
and $m_{0}=0.0115$ respectively, while the crosses
represent the rescaled one for $L_1=256$ with $m_{01}=0.0050$.
The solid line
fitted to the crosses is the persistence probability
for $L_2=128$ with $m_{02}=0.0101(1)$.
From the figure, we see clearly the data collaps nicely
and we can conclude that 
the scaling form in Eq. (\ref {e40}) holds.
The estimated scaling dimension for $m_0$ is
$x_0=1.01(1)$. To confirm this result,
we have also done similar simulations for
$L_1=256$ with $m_{01}=0.01$ and obtained
$x_0=1.06(4)$. Both values are consistent and 
the finite $m_0$ effect is already quite small.
Further, the simulations with the Metropolis algorithm show
similar results.
However, the value of $x_0$ measured here
is very different from that measured from
the time evolution of the magnetization or the auto-correlation,
 $x_0=0.536(2)$ \cite {oka97a}. This fact
 remains to be understood.

In conclusion, we have numerically measured 
the global persistence probability $p(t)$.
The critical exponent $\theta_1$ is directly
determined for the two--dimensional Ising and Potts model
with both the heat-bath and the Metropolis algorithm.
Within the statistical errors, universality is confirmed.
Compared with the previous results for the Ising model
with the heat-bath algorithm, our simulations have been
carried out with special attention
to the dependence of the scaling behaviour 
on the initial magnetization.


\begin{thebibliography}{10}

\bibitem{jan89}
{H. K. Janssen, B. Schaub and B. Schmittmann}, Z. Phys. {\bf {B 73}} (1989)
  539.

\bibitem{hus89}
D.~A. Huse, Phys. Rev. {\bf {B 40}} (1989) 304.

\bibitem{hum91}
K. Humayun and A.~J. Bray, J. Phys. A {\bf {24}} (1991) 1915.

\bibitem{li94}
{Z.B. Li, U. Ritschel and B. Zheng}, J. Phys. A: Math. Gen. {\bf {27}} (1994)
  L837.

\bibitem{gra95}
P. Grassberger, Physica {\bf {A 214}} (1995) 547.

\bibitem{sch95}
L. {Sch\"ulke} and B. Zheng, Phys. Lett. {\bf {A 204}} (1995) 295.

\bibitem{oka97a}
{K. Okano, L. {Sch\"ulke}, K. Yamagishi and B. Zheng}, Nucl. Phys. {\bf {B
  485}} (1997) 727.

\bibitem{oka97}
{K. Okano, L. {Sch\"ulke}, K. Yamagishi and B. Zheng}, {\em Monte Carlo
  simulation of the short-time behaviour of the dynamic XY model.}, Siegen
  Univ., 1997, preprint SI-97-02.

\bibitem{maj96a}
{S.N. Majumdar, A.J. Bray, S. Cornell and C. Sir}, Phys. Rev. Lett. {\bf {77}}
  (1996) 3704.

\bibitem{sta96}
{D. Stauffer}, Int. J. Mod. Phys. {\bf {C7}} (1996) 753.

\bibitem{oer97}
{K. Oerding, S.J. Cornell and A.J. Bray }, {\em Non-Markovian Persistence and
  Nonequilibrium Critical Dynamics}, {Univ. D\"usseldorf and Univ. Manchester},
  1997, preprint.

\bibitem{tan87}
S. Tang and D.~P. Landau, Phys. Rev. {\bf {B 36}} (1987) 567.

\bibitem{lan88}
{D. P. Landau, S. Tang and S. Wansleben.}, Journal de Physique Colloque {\bf
  49} (1988) C8.

\bibitem{ito93}
N. Ito, Physica {\bf {A 196}} (1993) 591.

\bibitem{li95}
{Z.B. Li, L. {Sch\"ulke} and B. Zheng}, Phys. Rev. Lett. {\bf {74}} (1995)
  3396.

\bibitem{li96}
{Z.B. Li, L. {Sch\"ulke} and B. Zheng}, Phys. Rev. {\bf E 53} (1996) 2940.

\bibitem{nig96}
{M.P. Nightingale and H.W.J. Bl\"ote }, Phys. Rev. Lett. {\bf {76}} (1996)
  4548.

\end{thebibliography}


\begin{table}[h]\centering
$$
\begin{array}{|c|l|l|l|}
\hline
   L    &   \multicolumn{3}{c|} {256} \\ 
\hline
  m_0  & random & .0010 & .0005 \\
\hline
\theta_1 & .233(5) &  .237(5) & .238(3) \\
\hline
\end{array}
\quad
\begin{array}{|c|l|}
\hline
    L    &   \multicolumn{1}{c|} {512}\\
\hline
  m_0   &  .0005  \\
\hline
\theta_1 & .238(6)\\
\hline
\end{array}
$$
\caption{
 The exponent $\theta_1$ measured for the two--dimensional Ising 
model
with the heat-bath algorithm. `Random' represents the result
obtained without the sharp preparation technique.
}
\label{t1}
\end{table}

\begin{table}[h]\centering
$$
\begin{array}{|c|l|l|l|l|}
\hline
       &   \multicolumn{2}{c|} {Ising}   & \multicolumn{2}{c|} 
{Potts}
\\
\hline
    &\quad  HB &\quad ME &\quad HB &\quad ME\\
\hline
\theta_1 & .238(3) &  .236(3) & .350(1) & .350(8)\\
\hline
\alpha_1 & .215(1) &  .212(2) & .320(3) & .324(3)\\
\hline
\end{array}
$$
\caption{
 The exponent $\theta_1$ measured for the two--dimensional Ising 
and 
Potts model with both the heat-bath (HB) and the Metropolis (ME) 
algorithm.
The initial magnetization is $m_0=0.0005$ and the lattice sizes are 
$L=256$
and $288$ for the Ising and Potts model respectively.
The exponent $\alpha_1 = - \theta + (1-\beta/\nu)/z$ is
taken from reference [7].
}
\label{t2}
\end{table}

\newpage

\end{document}